\documentclass{llncs}
\usepackage{graphicx} % Required for inserting images
\usepackage{cite}
\usepackage{xcolor}
\usepackage{amssymb}
\usepackage{algorithm}
\usepackage{array}
\usepackage{multirow}
\usepackage[
  separate-uncertainty = true,
  multi-part-units = repeat
]{siunitx}
\usepackage{algpseudocode}
\begin{document}
\title{MUSE-Explainer: Counterfactual Explanations for Symbolic Music Graph Classification Models}

\newcommand{\mk}[1]{\textcolor{orange}{(m) #1}}

\author{Baptiste Hilaire\inst{1,2}, Emmanouil Karystinaios\inst{2}, \and Gerhard Widmer~\inst{2}}

\institute{ENS Paris-Saclay, Gif-sur-Yvette, France \and Institute of Computational Perception, Johannes Kepler University Linz, Austria \\
\email{firstname.lastname@jku.at}}

\maketitle

\begin{abstract}

Interpretability is essential for deploying deep learning models in symbolic music analysis, yet most research emphasizes model performance over explanation. To address this, we introduce MUSE-Explainer, a new method that helps reveal how music Graph Neural Network models make decisions by providing clear, human-friendly explanations. Our approach generates counterfactual explanations by making small, meaningful changes to musical score graphs that alter a model’s prediction while ensuring the results remain musically coherent. Unlike existing methods, MUSE-Explainer tailors its explanations to the structure of musical data and avoids unrealistic or confusing outputs. We evaluate our method on a music analysis task and show it offers intuitive insights that can be visualized with standard music tools such as Verovio. All code and models are available on \url{github.com/BaptisteHi/MNExplainer}.

    %Explainability in AI is still expanding in many fields. But in Music Information Research, there remains a lack of explainability tools that are tailored for giving insights on symbolic music data analysis models prediction. We introduce MUsical noiSE (MUSE)-Explainer, a tool designed for producing counterfactual explanations for GNN classification models performing on symbolic music data. Inspired by the concept of noise diffusion explainers, MUSE-Explainer makes subtle changes to the structure of the musical data, avoiding unrealistic or confusing outputs.
    %% that would make sense musically, such as adding a note in the score or modifying a note's pitch, in order to flip the model's prediction to the desired one. 
    %We evaluate our method on a music analysis task and show it offers intuitive insights that can be visualized with standard music tools such as Verovio. 
    %% We then provide a scoring system for these counterfactual explanations produced, and investigate what changes are the most successful at changing the model's outcome and at giving musically intuitive insights.

\end{abstract}

\section{Introduction}

In the past years, the field of symbolic music processing has experienced a boom in deep learning techniques that have provided high-performing solutions for many analysis tasks~\cite{sailorrnbert}, music generation~\cite{chen2024sympac}, emotion recognition~\cite{bhuvana2023emotion}, expressive performance~\cite{jeong2019virtuosonet}, and composer classification~\cite{zhang2023symbolic}. 
While deep learning has proven effective for symbolic music processing, deploying these models in practice also requires understanding their predictions, which poses a challenge in particular for big black-box models.

Focusing on music analysis, a class of deep learning models, called Graph Neural Networks, has shown promising performance at capturing complex relationships in music data~\cite{graphexprperf,  graphmuse}. Despite the importance of interpretability in symbolic music processing and analysis in particular, most research still focuses on improving model performance rather than explaining how these models work. The few existing methods~\cite{smugexplain} for explaining graph neural networks for music provided limit insight with general models not tailored for music.

% Introduction de paragraph
Graph-based models are good candidates for explainability in music because they can capture interrelations between notes and provide insights at the level of individual note instances. Model-agnostic explainers are considered more general because they do not rely on the gradients of the prediction model. Many of those explainers rely on perturbing an input to explain the predictions. Sometimes, the perturbation process can result to out-of-distribution issues, that just increase the instability of the underlying prediction model. To tackle the particularities of out-of-distribution issues, we designed a novel explainer based on graph-based counterfactual explanations, MUSE-Explainer. 

Counterfactual explanations are considered human-friendly because they focus on changing small parts of the input in order to obtain a desired outcome~\cite{molnar2025}. Our explainer, MUSE-Explainer, produces counterfactual explanations for model predictions on musical score graphs. A musical score can be modeled by an oriented heterogeneous graph, which is a graph where each node is associated with a node type, and each edge with an edge type~\cite{graphexprperf}. MUSE-Explainer is capable of dealing with out-of-distribution issues by ensuring the musical coherence of the score represented by the counterfactual explanation graphs it produces.

% Counterfactual explanations are preferred for their intuitive aspect \cite{molnar2025}, easing the process of retrieving information about how is a model making its predictions, 

% Counterfactual explanations are human-friendly explanations of a target prediction, since they focus on small aspect of the input and propose a concrete way of changing it to obtain a desired outcome. They consist of a modified version of the input for the explained prediction that is as close as possible to the original, while affecting consequently the model output. For instance, if the model is a classification model, a counterfactual explanation should flip the predicted label of the original input it was obtained from.

Our contributions in this work are four-fold:
\begin{itemize}
    \item An explainer for Heterogeneous GNNs classification models performing on node level, that produces musically intuitive counterfactual explanations
    \item Five musical edit operations to modify an input graph in sensible and interpretable ways
    \item A configuration study of our explainer on a cadence detection model to measure its accuracy and efficiency
    \item A compatible output with visualizing frameworks that provides a view on the edited musical score in Verovio.
\end{itemize}

% In this work, we contribute MUSE-Explainer, a counterfactual explainer based on a new musically intuitive approach that permits the generation of a desired amount of counterfactual explanations which are obtained from an increasing number of changes applied to the provided input graph. We then provide a rating system of the obtained counterfactual explanations, and help visualizing them using the SMUG-Explain framework \cite{smugexplain}.  We will focus on explaining classification GNN models since many symbolic analysis tasks are handled by those \cite{cadence, voicesep}. With this work, we pave the way for some more music-intuitive interpretability counterfactual methods. These would provide some appealing insights of how is a symbolic music model performing, ultimately allowing us to retrieve some musical knowledge from the functioning of our models.

\section{Related work}

\subsection{Explainability for Graph Neural Networks}

In the past years, geometric deep learning has gained traction, leading to a wide array of models and research focused on graph-structured data. As a result, various approaches to explainability have been developed specifically for Graph Neural Networks (GNNs). These approaches are usually categorized in a taxonomy based on their underlying explanation strategies, such as model agnostic vs model-aware modes but also methods such as gradient-based, perturbation-based, surrogate models, and counterfactuals, each providing different perspectives on model interpretability.

The model-aware methods are specific to a model, treating it as a white box, and accessing its parameters. For instance, the integrated gradients method~\cite{integratedgradients} proposes to use the gradient to retrieve the input features and a subgraph of the input graph that played a major role in the prediction. Saliency maps~\cite{saliencymaps} also offer some insights about feature importance in the prediction. While these methods are often effective, their reliance on detailed model internals limits their flexibility and prevents them from being easily applied to different model architectures.

In contrast, model-agnostic methods treat the model as a black box, relying solely on its inputs and outputs. For example, LIME~\cite{lime} builds an interpretable surrogate model around a specific prediction by generating many slight perturbations of the input and observing the resulting outputs, thus offering local explanations of the black-box model’s behavior. Another method, GNNExplainer~\cite{gnnexplainer}, a perturbation-based method, learns an edge mask to identify the subgraph most relevant to a model’s prediction, highlighting the key structures that influenced the outcome. However, both approaches have their limitations. With LIME, small changes to the input can sometimes yield surrogate models whose explanations vary significantly, reducing the reliability of the interpretation. Meanwhile, GNNExplainer is prone to out-of-distribution issues, as the masked subgraphs it produces may not accurately reflect the kinds of data the model was originally trained on, potentially leading to unstable explanations.

%Quelques doutes sur ce paragraphe, mais il faut introduire les counterfactual ici
Although both model-aware and model-agnostic methods contribute valuable explanations, they can struggle with reliability and stability, especially when input modifications produce unrealistic or out-of-distribution samples. An alternative, increasingly popular direction is to use counterfactual explanations, which provide human-friendly insights by illustrating how small changes to the input can alter the model’s prediction~\cite{CFGNN, d4explainer}, all while treating the model as a black box. Depending on the algorithm used to generate them, counterfactual methods can help avoid out-of-distribution issues, a property we demonstrate in our work for music graph classification tasks.

% Beyond explainability, graph-based methods have been applied in symbolic music analysis and manipulation, including work by Miller and Sandler~\cite{MillerSandler}, Metzig and Sandler~\cite{MetzigSandler}, and Nardelli et al.~\cite{Nardelli}, which highlights the broader relevance of graph representations for musical data.

%Along the increasing development of deep learning models performing on graphs came the need to understand these models and get some insight about their inner workings. There are many ways of retrieving some information about these mechanisms, hence many branches of explainability in AI that can be applied to Graph Neural Networks. An overview of the field is proposed in an online book written by Molnar \cite{molnar2025}.

%Some methods, classified as model-specific methods, make deductions out of the parameters of the target model, looking at what part of an input will cause most of neuron activation for instance \cite{saliencymaps}.

%The counterpart of these methods are model agnostic techniques where the model is treated as a black box, and deductions on the model inner workings can be made by manipulating the inputs provided to the model. Among these methods, counterfactual explanations emerged as a human friendly way of understanding what aspect of an input can affect the prediction of the model. Existing work proposed generating counterfactual explanations for GNNs \cite{d4explainer, CFGNN}, but these methods are not suited for our very specific data structure, causing out of distributions issues.
\subsection{Music Information Research}

In the field of Music Information Research, existing work aims to explain deep learning models and assess the trustworthiness of different explainers. In the symbolic music domain, Foscarin et al.\cite{DBLP:conf/ismir/FoscarinHPFW22} introduce a method to identify which musical concepts are relevant to a given prediction. In the audio domain, Hoedt et al.\cite{DBLP:journals/nca/HoedtPFW23} investigate various explainers using adversarial attacks to determine whether these methods can reliably identify the parts of the input that are crucial for the model’s prediction.

Focusing on graph neural networks for music, the SMUG-Explain framework~\cite{smugexplain} introduces visualization tools for explanations generated by standard gradient-based methods such as Integrated Gradients~\cite{integratedgradients} and Saliency maps~\cite{saliencymaps}. To our knowledge, this is the only explainability method specifically developed for symbolic music analysis using graph neural networks, highlighting the need for further research on interpretable solutions in this field.

\section{Methodology}

\subsection{Graph Structure}

This work is focused on producing explanations for graph-based neural networks for music analysis, therefore the input data, i.e. the musical score, is represented as a graph. We choose to use an established method in the field to construct the heterogeneous directed graphs from the scores as introduced in \cite{graphexprperf}. One node type 'note' is used to create a node for every note in the score. Then, four types of edges can be found in our graphs. Let us note $\mathbf{on}(u)$ and $\mathbf{dur}(u)$, respectively the onset and the duration of a note node $u$, and let us consider two note nodes $u$ and $v$. 

\begin{enumerate}
    \item If $\mathbf{on}(u) = \mathbf{on}(v)$, then an 'onset' edge is created from $u$ to $v$.
    \item If $\mathbf{on}(u) + \mathbf{dur}(u) = \mathbf{on}(v)$, a 'consecutive' edge is created from $u$ to $v$.
    \item If $\mathbf{on}(u) < \mathbf{on}(v) \land \mathbf{on}(u) + \mathbf{dur}(u) > \mathbf{on}(v)$, then it is a 'during' edge that is added from $v$ to $u$.
    \item Finally, if $\mathbf{on}(u) + \mathbf{dur}(u) < \mathbf{on}(v) \land \nexists w \in V, \mathbf{on}(w) < \mathbf{on}(v) \land \mathbf{on}(u) + \mathbf{dur}(u) < \mathbf{on}(w)$, then a 'rest' edge is created from $u$ to $v$, where $V$ is the set of note nodes.
\end{enumerate}

Some other nodes can be added to the graph structure using the process introduced in Graphmuse~\cite{graphmuse}, such as 'beat' and 'measure' nodes, to connect the nodes representing notes to the beat or measure they belong to in the musical score. This hierarchical information can be beneficial by allowing the graph neural networks to access more complex relations in the score, leading to better results when resolving tasks that are tied to complex musical structures~\cite{graphmuse}. %J'aimerai pouvoir affirmer ceci, donnant une motivation à cet ajout potentiel, mais cet ajout a t'il vraiment aider pour des modèles comme la détection de cadences ? Sinon ce paragraphe entier peut aussi être retiré simplement.

\subsection{Algorithm and Workflow}

As noise diffusion processes proved their efficiency to generate counterfactual explanations~\cite{d4explainer}, we create our own algorithm inspired by them, adding up an increasing number of subtle changes to the graph encapsulating the musical score. Similar to noise diffusion, we start with an input graph and progressively create noisier versions. However, instead of introducing random noise, we apply musically coherent modifications, ensuring that the resulting graphs remain within the input distribution. This approach eliminates the need for denoising or post-processing, allowing the noisy graphs to serve directly as valid counterfactual explanations while avoiding out-of-distribution issues.

\begin{figure}
    \centering
    \includegraphics[width=1\linewidth]{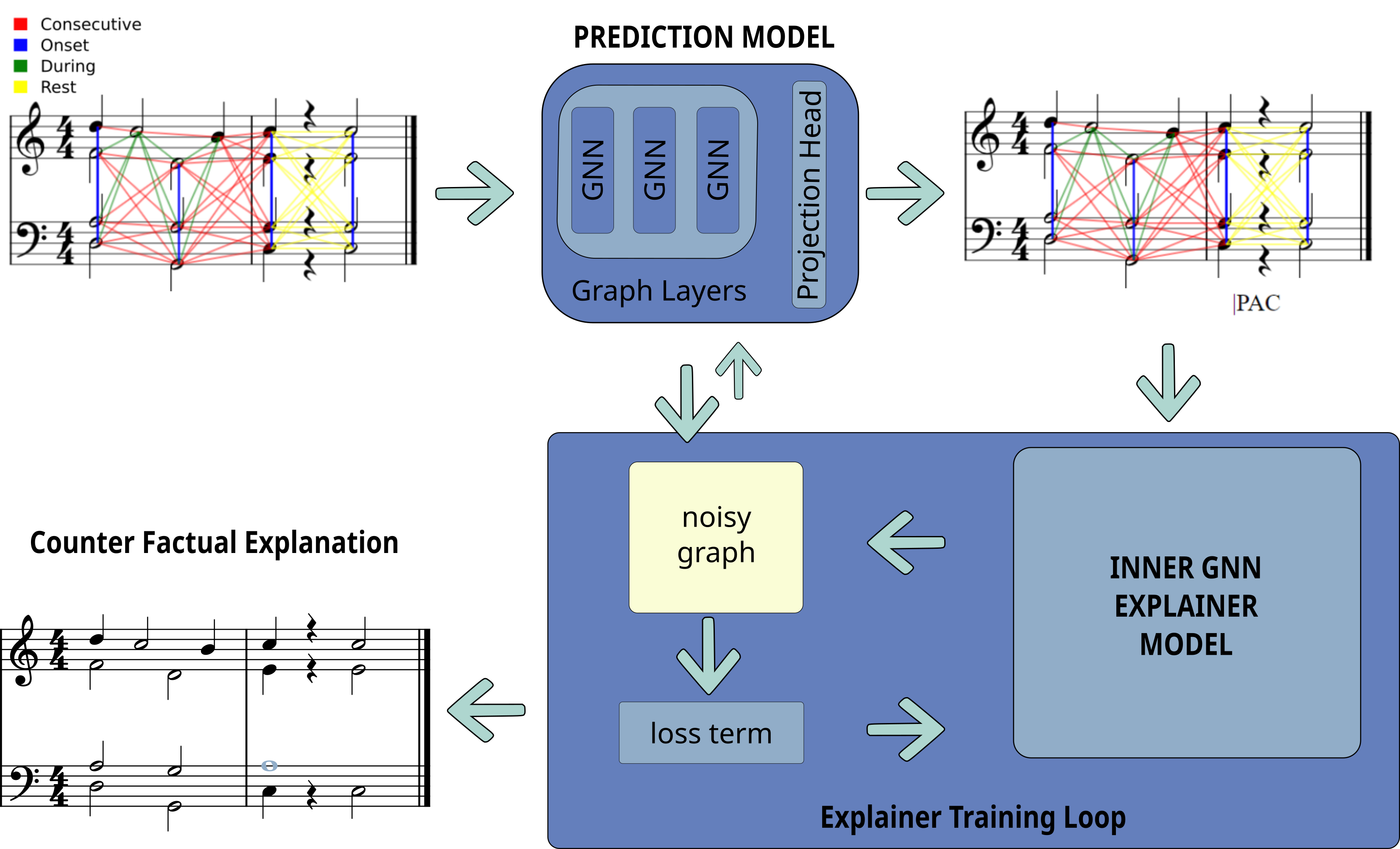}
    \caption{Overall workflow of MUSE-Explainer (input graph picture taken from \cite{cadence}).}
    \label{fig:workflow}
\end{figure}

The overall workflow of our explainer is illustrated in Figure~\ref{fig:workflow}, and the algorithmic details are provided in Algorithm~\ref{alg:cap}. Our explainer is built with two parts : an outer component that receives the parameters such as the number of training epochs and the input graph, and an inner model that is used to learn the changes that could flip the label predictions of the explained model. When created, he explainer is initialized with a selected classification model. For each explanation to be generated, an input graph and a target are provided to the explainer’s outer component. This component then enters a training phase, using the inner model to learn both the type of operation to apply and its associated parameters. At the end of training, a counterfactual explanation is produced, and a loss score is computed that considers both the distance from the input graph and the validity of the counterfactual. The explainer then backpropagates this loss to the inner model, which refines the choice of operation and its parameters. By repeating this training step, the explainer iteratively generates a sequence of counterfactuals, where each explanation is based on the modifications applied in the previous step, plus an additional learned change from the inner model.

\subsection{Training of the Inner Model and Loss Function}

The core component of the explainer is its inner model, which explores changes that affect the prediction and constructs the resulting explanations. To train this model effectively, we design a custom loss function that captures two essential properties of a high-quality counterfactual explanation: its counterfactual nature—specifically, i.e. the ability to flip the classification label, and its minimal distance from the original input data. The explicit form of this loss function is provided in Equation~\ref{eq:loss}.

\begin{equation}\label{eq:loss}
    \mathbf{L}(\mathbf{G},\mathcal{G},y,\lambda_{\mathbf{gp}},\lambda_{\mathbf{nd}}, \lambda) = \lambda \; \mathbf{ent}(f(\mathbf{G}),y) + (\lambda_{\mathbf{nd}} \mathbf{D}_{\mathbf{nd}}(\mathbf{G},\mathcal{G}) + \lambda_{\mathbf{gp}} \mathbf{D}_{\mathbf{gp}}(\mathbf{G},\mathcal{G}))    
\end{equation}

where $\mathbf{G}$ is the counterfactual graph, $\mathcal{G}$ is the original input graph, $y$ is the desired label output, $f$ is the model to be explained, $\lambda, \lambda_{\mathbf{gp}}$ and $\lambda_{\mathbf{nd}}$ are balancing factors and $\mathbf{ent}$ is a cross entropy function. $\mathbf{D}_{\mathbf{nd}}$ is a function adding up the the differences between the node features for each node that is present in both provided graphs, that also adds up the features from the nodes that are not shared between the graphs, such as added or removed notes. $\mathbf{D}_{\mathbf{gp}}$ is the standard graph edit distance, that can be computed efficiently by retrieving the possibly multiple changes that occurred while performing the operations to obtain $\mathbf{G}$ from $\mathcal{G}$, keeping track of the added and removed edges during the process.

\begin{algorithm}
\begin{algorithmic}
\Require $n > 0$ where $n$ is the number of explanations to produce, $\mathcal{G}$ the input graph, $t > 0$ the number of training steps.
\Ensure $\mathbf{L}$ contains the input graph and the $n$ explanations.
\State $\mathbf{L} \gets \mathbf{list}(\mathcal{G})$
\While{$n > 0$}
    \State $\mathbf{G} \gets \mathbf{L.last}$
    \For{$i \in \{1\dots t\}$}
        \State $\mathbf{G'} \gets \mathbf{InnerModel}(\mathbf{G})$
        \State $\mathbf{loss\_term} \gets \mathbf{loss\_function}(\mathbf{G'},\mathcal{G})$
        \State $\mathbf{loss\_term}.\mathbf{backward}$
    \EndFor
    \State $\mathbf{L.append}(\mathbf{InnerModel}(\mathbf{G}))$
\EndWhile
\end{algorithmic}
\caption{Production of counterfactual explanations for unspecified operations.}
\label{alg:cap}
\end{algorithm}

\subsection{Out of Distribution Issues}

Out-of-distribution issues typically occur when an input graph is modified by an explainer, for instance, from a noise diffusion model. These issues can be caused by introducing unseen feature values, drastic changes in edge distributions, graph size, or other structural shifts. In the context of music graphs, out-of-distribution problems are particularly pronounced due to the fundamental role of features and edges in encoding musical information. Every change to a music graph must correspond to a musically meaningful transformation in the underlying data. For example, if two note nodes $u$ and $v$ are already connected by an 'onset' edge, adding an additional 'consecutive', 'rest', or 'during' edge between them would introduce contradictions regarding their onset or duration, since these edge types have specific musical interpretations. Therefore, it is crucial to preserve musical coherence by carefully designing the operations that modify these graphs.

To address this, we first considered which kinds of modifications are sensible for a musical score. We propose five musically meaningful edit operations: updating pitch, onset, or duration, and adding or removing a note. The pitch update selects a 'note' node and adjusts its MIDI pitch, as well as its pitch spelling if supported by the input graph. Similarly, onset and duration updates involve selecting a 'note' node and changing its onset or duration. For onset updates, to avoid out-of-distribution issues, the new onset is set to that of another selected 'note' node, ensuring plausibility. Duration can be adjusted to standard musical values such as whole, half, quarter, or eighth notes. Removing a 'note' node entails disconnecting it from the graph by deleting all its edges. Adding a note involves inserting a new 'note' node with its pitch, onset, and duration determined in the same manner as in the update operations. By restricting changes to these carefully crafted operations, we ensure that the resulting graphs remain within the valid distribution of music graphs representing structured musical scores. As a result, our counterfactual explanations maintain musical coherence and do not suffer from out-of-distribution problems.

\section{Experiments and Results}

For our explainer, we conducted experiments using a cadence detection model originally introduced in~\cite{cadence} and further adapted in~\cite{graphmuse}. This model operates as a node-level classification Graph Neural Network (GNN), where the task is to predict whether a given node marks the occurrence of a musical cadence. The model leverages the structure of the musical score, encoded as a graph, to capture complex musical relationships and dependencies that inform its predictions. By applying our counterfactual explainer to this setting, we are able to assess both the interpretability and musical validity of our explanations in the context of a real symbolic music analysis task. We evaluate two scenarios: flipping from a \textbf{PAC} (Perfect Authentic Cadence) prediction to \textbf{NC} (No Cadence), and the inverse.

\begin{table}[!b]
    \centering
    \begin{tabular}{ p{1.5cm}||p{1cm}|p{1cm}|p{1.1cm}|p{1cm}|p{1cm}|p{1cm}|p{1cm}|p{1cm}}
         % \hline
         %  Results & \multicolumn{9}{|c|}{Explainer hyperparameters} \\         
         label & \multicolumn{4}{|c|}{PAC to NC} & \multicolumn{4}{|c}{NC to PAC} \\
         \hline
         balance & \multicolumn{2}{|c|}{Distance}& \multicolumn{2}{|c|}{Counterfactual} & \multicolumn{2}{|c|}{Distance}& \multicolumn{2}{|c|}{Counterfactual} \\
         \hline
         epochs & 50 & 100 & 50 & 100 & 50 & 100 & 50 & 100\\
         \hline 
         accuracy & 92.0\% & $\mathbf{100.0}$\% & 96.0\% & 92.0 \% & 20.0 \% & 16.0 \% & 8.0\% & $\mathbf{32.0}\%$  \\
         \hline
         min. changes & $1.4 \pm 1.1$ & $1.9\pm1.8$ & $2.3\pm2.6$ & $1.8 \pm 2.5$ & $1.6 \pm 0.8$ & $2.8 \pm 1.5$ & $5.0 \pm 1.0$ & $3.4 \pm 2.5$ \\
         \hline
         operation & onset & dur & onset & onset & rem & rem & dur & add  \\
         \hline
         distance & $2.8\pm2.2$ & $3.0\pm2.1$ & $0.3\pm0.2$ & $0.3 \pm 0.2$ & $4.0 \pm1.6$ & $3.1 \pm 1.9$ & $0.5 \pm 0.1$ & $0.3 \pm 0.2$  \\
         \hline
         time (s) & $68.5\pm 9.3$ & $134.7\pm 16.4$ & $66.8 \pm 9.9$ & $135.8 \pm 17.1$ & $68.5 \pm 10.4$ & $134.2\pm16.4$ & $67.5 \pm 9.6$ & $132.4 \pm 17.5$ \\
         \hline
    \end{tabular}
    \vspace{0.2cm}
    \caption{Results of the experiments led on five Mozart piano sonatas. The first column refers to what information is contained in the row : label is PAC to NC when we desire a no cadence label for a prediction that predicted a perfect cadence on the input graph. The balance refers to the loss function balancing term, distance is for $\lambda_{\mathbf{nd}} = \lambda_{\mathbf{gp}} = 0.1$ and counterfactual for $\lambda_{\mathbf{nd}} = \lambda_{\mathbf{gp}} = 0.01$, having $\lambda = 2.0$ for both settings. The accuracy is the accuracy of the explainer to flip the label for all of the tests. The min. changes refers to the number of operations needed to flip the label. The operation row provides the operation that succeeded the most frequently to flip the label, 'dur' standing for duration and 'rem' for the removal of a note. The distance is the loss function distance term between the counterfactual explanation that flipped the label and the input graph. Finally, the time in seconds is the production time for 10 counterfactual explanations built on top of each other. For each desired label setting, the best accuracy score is set in bold.}
    \label{tb:experiments}
\end{table}

% Content here:
Each experiment consists in producing ten counterfactual explanations for the prediction of a given piece and with given parameters such as balance factors, node target, and desired label prediction. Each experiment is repeated five times. We used five pieces for these experiments, randomly chosen in the Mozart piano sonatas dataset introduced by~\cite{dcml}. 
 
% These consist in the piano sonatas number 2, K.280, number 2, K.309, number 1, K.331, number 3, K.533 and number 2, K.545.

By analyzing Table~\ref{tb:experiments} we observe that the explainer is indeed able to produce counterfactuals. We highlight that the ground-truth label distribution is highly unbalanced: most notes are not associated with cadences. As a result, the model tends to predict NC more frequently, which makes generating PAC counterfactuals more challenging and partly explains the asymmetry in our results.

Despite this, our results demonstrate that the explainer succeeds in finding counterfactuals for those case, however, we remark that these changes require, on average, more operations and are slightly less accurate compared to the inverse process, which is coherent with the label distribution in the data.

Similarly, changing the number of training epochs for the explainer's inner model training phase had little impact on the results. The same observation holds for varying the balance between the counterfactual and distance loss terms.

% Modifying the balancing factors to privilege the term ensuring distance minimality or the term ensuring the counterfactual aspect of the explanation had less impact than expected. With counterfactual aspect being more accentuated, we had expected better results for accuracy, which was the case for some settings only, the inverse scenario occurring for others.

% In the end, the explainer provided great results for tasks made simpler by the label distribution in the data, managing to flip the labels consistently using a low amount of changes and while being trained only a few epochs. The results were less satisfying for harder label changes, but some successful explanations could still be produced.

\section{Discussion}

\subsection{Visualization and Scoring of the Explanations}

To facilitate interpretation and analysis, the explanations produced by our explainer can be visualized using the SMUG-Explain framework~\cite{smugexplain}, which is built on top of a Verovio-based browser interface. Within this framework, it is possible to directly observe the modifications made to the input graph, providing a clear and interactive view of the changes that led to the counterfactual explanation. In addition to graphical visualization, SMUG-Explain allows users to play back the musical score, so the effects of the modifications can also be experienced audibly. 

%rajouter intro
% The explanations produced by our explainer can be visualized using the SMUG-Explain framework~\cite{smugexplain}, giving access to a direct visual of the changes performed on the input graph. The framework can also play the score, so the changes performed can be heard.

Additionally, we propose a basic scoring method of the produced counterfactual explanations, measuring their distance from the original graph, their ability to flip the classification label to the desired one, and trying to retrieve the musical insights provided. This last aspect is the most interesting and the most challenging, since some modifications will provide clear insights, such as feature changes, and it will be much harder for others to get any precise understanding of the prediction process. In fact, the answer of what knowledge could be retrieved from the fact that adding a note at a given onset will change a model's prediction depends on many factors : it could either break a chord, or bring a feature value to this onset that would change the prediction, there could be many more options.

\subsection{A Sneak Pick in the Counterfactuals}

\begin{figure}
    \centering
    \includegraphics[width=1\linewidth]{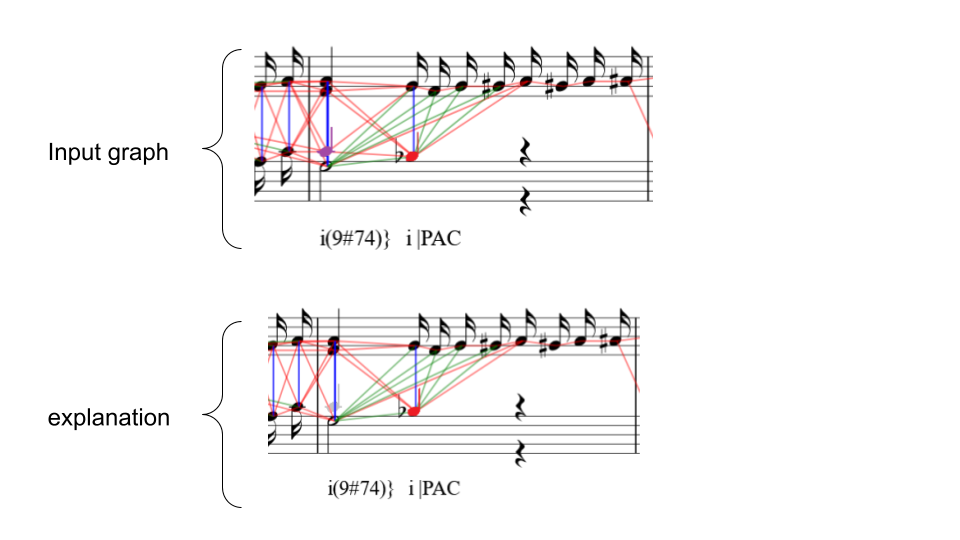}
    \caption{Visualization of a 'remove' operation performed by MUSE-Explainer on the purple note to remove a PAC label prediction on the red note.}
    \label{fig:ex}
\end{figure}

In this section, we demonstrate an example of the MUSE-Explainer attempting to interpret a GNN-based cadence detection model. In figure~\ref{fig:ex}, we showcase how the explainer changes the score to confuse the model and change its prediction.

We observe that the underlying model predicts a PAC cadence on the second beat of the measure that is entirely shown. The model's prediction is a result of a dominant resolution to tonic followed by a 9,7,4 suspension, and it is classified as a Perfect Authentic Cadence on the resolution of the suspension. Our explainer succeeds in flipping the models by removing on the suspension notes, which can be seen by its isolation in the graph (marked in light grey). While some modifications, such as removing a suspension note, provide clear harmonic insights, others, such as adding a new note, are harder to interpret consistently, since their effect depends on broader harmonic and structural context.

\subsection{System Modularity}

The MUSE-Explainer is designed with a modular architecture, enabling manual adjustment of all key parameters to suit specific tasks or research needs. When creating an instance of the explainer, users can configure the balancing factors $\lambda$, $\lambda_{\mathbf{nd}}$, and $\lambda_{\mathbf{gp}}$ in the loss function~\ref{eq:loss}, set the number of training epochs for the inner model, specify the number of convolutional layers in the encoders, and select the learning rate for model optimization.

Further customization is possible during the generation of counterfactual explanations. Users can choose the target label for the counterfactual prediction, tailoring the explanations toward specific classification outcomes. Additionally, the explainer supports the option to constrain the inner model to follow a user-defined sequence of operations when constructing counterfactuals. For example, one can prescribe a specific operation path—such as first performing a pitch change, followed by adding a new note—resulting in a sequence of counterfactual explanations, each corresponding to an incremental modification. This flexibility makes MUSE-Explainer adaptable to a wide range of musical analysis scenarios and user preferences.
% \subsection{System modularity}

% The \textbf{MUSE-Explainer} follows a modular design, which allows the manual adjustment of all its parameters. Starting with how an instance of the explainer is created, it is possible to adjust the balancing factors $\lambda, \lambda_{\mathbf{nd}}$ and $\lambda_{\mathbf{gp}}$ in the loss function \ref{eq:loss}, the number of training epochs when training the inner model for producing the explanations, the number of convolution layers in the encoders in this inner model, or the learning rate of the inner model.

% A lot of parameters can also be adjusted when producing counterfactual explanations for a prediction. In fact, the choice of the desired label for the classification of the counterfactual explanations is left open. There is also the possibility of forcing the inner model to choose a provided operation path to obtain the counterfactual explanations, which means it will learn the parameters for operations that are fixed from the start. Such path would be for instance a pitch change, then adding a new note, leading to the production of two counterfactual explanations obtained respectively with one and two changes.

\section{Conclusion and Future Work}

In this paper, we introduced a novel explainer that generates counterfactual explanations for classification graph neural networks operating on music graph data. To our knowledge, MUSE-Explainer is the first counterfactual explainer designed for symbolic music GNNs. By ensuring that modifications remain musically meaningful, it addresses common out-of-distribution issues and provides interpretable insights into model predictions.

For future work, we plan to extend our experiments to a broader range of graph neural network architectures, including edge-level and graph-level classification tasks. These scenarios may pose additional challenges, as modifying larger portions of the input graph may be necessary to alter the model’s prediction. Furthermore, we aim to enhance the efficiency of our explainer, enabling it to generate counterfactual explanations in parallel for multiple input graphs within the same model. This improvement would facilitate faster analysis and broader applicability in real-world symbolic music processing tasks.

\section*{Acknowledgements}
This work was supported by the European Research Council (ERC) under Horizon 2020 grant \#101019375 “Whither Music?”.

\bibliographystyle{splncs04}
\bibliography{biblio.bib}

\end{document}